# The screening effects on field enhancement factor of zigzag graphene nanoribbon arrays: A first-principles study

Han Hu[a], Tzu-Chien Lin[a], Tsan-Chuen Leung[a,*] and Wan-Sheng Su[b,c,*]

[a]Department of Physics, National Chung Cheng University, Chia-Yi 62101, Taiwan

[b]Experimentation Division, National Taiwan Science Education Center, Taipei 11165, Taiwan

[c]Department of Electro-Optical Engineering, National Taipei University of Technology, Taipei 10608, Taiwan

E-mail: tcleungtw@gmail.com (T. C. Leung); wssu@mail.ntsec.gov.tw (W. S. Su)

Abstract

The field screening effect on the electronic and field-emission properties of zigzag graphene nanoribbons (ZGNRs) has been studied using first-principles calculations. We have systematically investigated the effect of inter-ribbon distance and ribbon width on the work function, field enhancement factor, band gap and edge magnetism of zigzag graphene nanoribbons (ZGNRs). It is found that the work function of ZGNRs increases rapidly as the inter-ribbon distance Dx increases, which is caused by the positive dipole at the edge of the ribbon. For a given Dx, the work function of ZGNRs decreases as the ribbon width W increases. The wider the width of ribbon, the stronger the effect of inter-ribbon distance on the work function. Using a simple linear interpolation model, we can obtain the work function of ZGNR of any ribbon-width. For the case of Dx < W, the field enhancement factor increases rapidly as the inter-ribbon distance increases. As we further increase Dx, the enhancement factor increases slowly and then tends toward saturation. The inter-ribbon distance of ZGNRs can modulate the magnitude of the band gap and edge magnetism. These observations above can all be explained by the screening effect.

Keywords: first-principles, screening effect, graphene nanoribbon, work function

## Introduction

The first two-dimensional (2D) material graphene, composed of a single atomic sheet of graphite, has been successfully synthesized in laboratory[1-6] and has ignited intense investigation. Its electrons and holes behave like a massless Dirac fermion. Graphene (2D), with its planar, hexagonal arrangement of carbon atoms, is the basic structure forming graphite (3D), carbon nanotubes (1D), and fullerenes (0D). These novel forms of carbon present unique opportunities to study low-dimensional physical phenomena. Motivated by interest in fundamental physics and by the recognition of their unique and versatile applications[7-9], graphene and related structures have been extensively investigated[10-13]. A quasi-one-dimensional graphene nanoribbon is constructed by cutting or by patterning a graphene sheet along a specific direction[5,6,14]. Such a graphene nanoribbon is essentially a strip of graphene with finite width, nanometers in size. Graphene nanoribbons with zigzag shaped edges are generally referred to as zigzag graphene nanoribbons (ZGNRs). The ZGNRs have unusual electronic localized edge states which decay exponentially into the center of the nanoribbon[14]. The edge states are a two-fold degenerate flat band at the Fermi level, with the flat band lasting about one-third of the Brillouin zone starting from the zone boundary. While counting the spin polarization effect, the ground state of ZGNRs appears as antiferromagnetic. That is, the two edges of the ribbon prefer to exhibit opposite spin orientations and the total spin is zero. Thus, the degenerated edge states split into two states with opposite spins and an open gap. The ZGNRs become semi-conducting but not metallic[15-19]. The band gap varies with the width of the ZGNR. Tuning the electronic properties of the ZGNRs has been a subject of interest to both experimental and theoretical researchers, with the motivation of finding possible applications. The ZGNRs may become half-metallic when an external electric field is applied across the nanoribbon, as suggested by Son *et al.*[16] The effect of this external electric field is to cause the band gap to close in one spin direction and to open in the other spin direction. Besides an external electric field, the distortions and defects of the ZGNR are also predicted to have an effect on band structures[20-21]. Furthermore, doping with various types of foreign atoms, adsorption of foreign atoms or molecules and external stress are other effective ways to widen their possible applications[22-24].

Owing to the structure being similar to carbon nanotubes, graphene nanoribbons are expected to have various unique properties and potentials for applications in nanoelectronics and nanomechanical devices. The high aspect ratio, excellent electrical and thermal conductivity, and good mechanical properties make graphene nanoribbon a promising candidate for use as an electron field emitter. Recent experimental results have demonstrated the possibility of using graphene nanoribbon as a field emitter[25-30]. Vertically aligned graphene nanoribbon arrays synthesized by chemical unzipping multiwall carbon nanotubes exhibits a low emission threshold field and remarkable current stability[31-32]. Work function and field enhancement factor are very important factors for field emission. Previous experimental and computer simulation studies show that the work function and the field enhancement factor of finite-length carbon nanotube arrays can be strongly affected by intertube distance[33-36]. When the length of the carbon nanotube is longer than the intertube distance, the field enhancement factor of the finite-length carbon nanotube arrays increases rapidly as the intertube distance increases. This phenomenon is caused by the field screening effect. The electric field penetration between carbon nanotubes (CNTs) increases as the intertube distance increases, leading to an increase in the field enhancement factor. Up to now, there has been no comprehensive systematic study of the effect of inter-ribbon distance and ribbon width on the work function, field enhancement factor, band gap and edge magnetism of zigzag graphene nanoribbon, and this will be the focus of this work.

Results and discussion

Figure 1(a) shows the atomic structures of a 6-ZGNR, with the coordinate axes defined in Fig. 1(b). We refer to a zigzag graphene nanoribbon with N zigzag chains as a N-ZGNR. The dangling bonds at the edges of the ribbon are passivated by hydrogen atoms. The nanoribbon edge direction is along the z-axis and a manipulated vacuum thickness, termed Dx, is applied to the x axis, with a vacuum space (Dv) of 20 Å or more along the z direction. The ribbon width (W) is defined here as the width without including the hydrogen atoms at the edges. The widths of the N-ZGNRs for N equal to 4, 6, 8, 10, 12 are 7.05, 11.29. 15.53, 19.76, and 24.01 $A^0$, respectively.

We first consider the effect of the ribbon width (W) and inter-ribbon distance (Dx) on the work function of zigzag graphene nanoribbons. The work functions were calculated by the difference between the Fermi level and the average potential in the vacuum region where it approaches a constant. For the case of a semiconducting ribbon, the Fermi level is chosen at the middle of the gap. Figure 2(a) shows the calculated work functions as a function of the ribbon width for various inter-ribbon distances. Our local-density approximation results agree with a previous study[37]. For a given inter-ribbon distance, the work function of ZGNR increases as the width of the ribbon decreases. It is of interest to note that the work function is strongly dependent on the inter-ribbon distance. This dependence can be qualitatively explained by the fact that there is a dipole at the edge of the ribbon. Carbon attract electron from hydrogen; therefore, the dipole at the edge is positive. A positive dipole layer leads to a decrease in work function. The higher the density of the dipoles, which corresponds to smaller inter-ribbon distance, the lower the work function. This explains the observation that the work function decreases as the inter-ribbon distance decreases.

We see from Fig. 2(a) that the work function varies roughly linearly with 1/W as W is greater than 15.53 A$^0$ ( corresponding to N > 8 ). This allows us to extrapolate to 1/W ➔ 0, which will give the work function of a "wide" ribbon. Figure 2(c) shows the work function of the wide ZGNR as a function of inter-ribbon distance calculated by the linear fit of the data in Fig. 2(a). It is of interest to note that the work function of the wide ZGNR also varies linearly with 1/Dx. The straight line in Fig. 2(c) is the linear fit to the data which is given by ϕ = 4.35 – 4.71/Dx. An extrapolation toward a larger Dx limit gives a value of 4.35 eV, which is the work function of an isolated wide ZGNR. The theoretical and experimental work functions of graphene are 4.5 eV and 4.6 eV, respectively[38-39]. Note that the work function of isolated wide ZGNR is lower than that of graphene. This is because the edge of the ZGNR is terminated by a hydrogen atom in our calculation. The positive dipole layer at the edge of the ribbon will lower the work function of the system. In order to confirm this observation, we plot the work function of ZGNR as function of inter-ribbon distance for various ribbon widths, as shown in Fig. 2(b). We find that the work function varies roughly linearly with 1/Dx as $D_x \geq 30$. The linear fit results will give the work function of an isolated ZGNR with various ribbon widths, as shown in Fig. 2(d). The work function of an isolated ZGNR is also linear, as given by ϕ = 4.35 – 0.62/W. An extrapolation to 1/W ➔ 0 will give the work function of an isolated wide ZGNR. The work function of an isolated wide ZGNR from Fig. 2(d) agrees well with that from Fig. 2(c). Therefore, the linear interpolation results in Fig. 2(a) and 2(b) are reliable. This means that we can use the results of linear regression in Fig. 2(a) to obtain the work function of a ZGNR of any ribbon-width.

We now study how the inter-ribbon distance affects the field enhancement factor of ZGNRs. In order to simulate a homogenous external electric field (E) in the z-direction (perpendicular to the edge) as shown in Fig. 1(b), we impose an electric field by applying a classical dipole sheet at the middle of the vacuum region in the z direction. The field enhancement factor ($\beta$) is defined as the ratio of the maximum value of the local electric field in the vicinity of the edge of the ZGNR to the applied field. The local electric field is determined by the gradient of the Coulomb potential energy difference induced by the applied field. Figure 3 shows the spatial distribution of field-induced coulomb potential around the 8-ZGNR under an external electric field of $0.1 V/A^0$ with inter-ribbon distances equal to 10 and 20 $A^0$. The images are viewed from the plane y = 0. We can see that the electrostatic field penetration increases as the inter-ribbon distance increases, which results in a larger field enhancement factor. The enhancement factors of 8-ZGNR under an external field of 0.1 V/ $A^0$ are 1.71 and 2.36 for Dx equal to 10 and 20 $A^0$, respectively. The enhancement factors are not sensitive to the strength of the external field as long as the field is not too large. For example, the enhancement factor of 8-ZGNR with Dx = 20 $A^0$ are 2.36, 2.36, 2.39, 2.56, and 2.70 for external fields of 0.05, 0.10, 0.15, 0.20, and 0.25 V/ $A^0$, respectively. Note that the maximum of the field is not located at the top of the edge, as shown in Fig. 3(a). In order to better understand these observations, we plot the field-induced Coulomb potential and field-induced charge density in the vicinity of the edge of a 8-ZGNR with Dx = 10 $A^0$, as in Fig. 3(c) and 3(d). It is found that the maximum electric field is located at ( -0.185, 0, 7.957 ) and ( 0.185, 0.0, 7.957) $A^0$, and the electric field at these two points is given by (-0.09, 0.0, -0.14) and (0.09, 0.0, -0.14) $V/A^0$, respectively. We can clearly see from Fig. 3(d) that the position of the field maximum is closely related to the field-induced charge density. The spatial distributions of field-induced Coulomb potential and field-induced charge density in the vicinity of the edge of ZGNR are almost independent of the ribbon-width and inter-ribbon distance.

Figure 4(a) shows the calculated field enhancement factors as a function of inter-ribbon distance for various ZGNRs under an electric field of 0.1V/ $A^0$. When the inter-ribbon distance is smaller than the width of the ribbon, the enhancement factor grows rapidly as Dx increases. When we further increases Dx, the enhancement factor increases slowly and then tends toward saturation. As the inter-ribbon distance becomes much larger than the width of the ribbon, the enhancement factor remains the same as that of an isolated ribbon. This phenomenon is caused by the electrostatic screening effect, which is determined by the width of the ribbons compared to the inter-ribbon distance. When the ribbons are too close together for electric field penetration, it will lead to the suppression of the field enhancement. The saturated value of the enhancement factor increases as the ribbon width increases. The dependence of the enhancement factor on the ribbon width W and inter-ribbon distance Dx can be well described by the following expression[40].

$$\beta = c_1 + c_2 \left[1 - \exp(-c_3(Dx))\right] \qquad (1)$$

The dashed line in Fig. 4(a) is the curve-fitting results. The fitting values of $c_3$ are given by 0.129, 0.092, 0.077, 0.048, and 0.036 (1/$A^0$) for N = 4, 6, 8, 10, and 12, respectively. In addition, the saturated β values for ZGNR of N = 4, 6, 8, 10, and 12 obtained from the fitting curve are 2.10, 2.37, 2.63, 3.20 and 3.93, respectively.

In order to describe the screening effect more quantitatively, we calculate the value of Dx such that the enhancement factor is 50, 60, 70, 80, 90 and 99% of its saturated value for various ribbon widths, using the curve-fitting results from Fig. 4(a). The results are shown in Fig. 4(b). The solid line in Fig. 4(b) is the third degree polynomial curve-fitting result. When the ribbon width is smaller than 20 $A^0$, the enhancement factor almost reaches 90% of its saturated value as Dx ➔ 2W. For the case of 12-ZGNR, the value of Dx which reaches 90% of its saturated enhancement factor is 2.5W. Therefore, the wider the ribbon, the larger the ratio of the inter-ribbon distance and ribbon width ( Dx/W ) which is needed to suppress the screening effect.

In the following we investigate how the dependence of the band gap and edge magnetism on external electric field is affected by the inter-ribbon distance. As revealed in previous studies[16], the ground state of the ZGNR is as an antiferromagnetic semiconductor with antiparallel spin orientation between the two edges. The states of opposite spin orientation are degenerate, and the band gap is inversely proportional to the width of the ribbon. When an external electric field is applied across the nanoribbon, the degeneration of the edge states of opposite orientation is split. One of the spin states, labeled as $\alpha$-spin, will wider its gap while the other spin state, labeled as $\beta$-spin, will close its gap. Figure 5 shows the calculated band gap of ZGNRs as a function of inter-ribbon distance under an external electric field of 0.0, 0.1 and 0.2 $V/A^0$. For the case of zero-applied electric field, the $\alpha$-spin and $\beta$-spin states are degenerate and the band gap is inversely proportional to the width of ribbon; this observation agrees with previous reports[17]. It is of interest to note that the inter-ribbon distance can modulate the magnitude of the band gap. The band gap of ZGNRs will increase as the inter-ribbon distance increases, showing a saturation tendency as the inter-ribbon distance is greater than the width of the ribbon. The wider the width of ribbon, the stronger the dependence of the band gap on the inter-ribbon distance. The physical origin of this phenomenon is due to the screening of the electric field generated by the edge dipole of the ribbon.

Next, we study the effect of inter-ribbon distance on the band gap of ZGNRs when an external electric field is applied. The external field induces splitting of the double degenerate edge states, therefore the band gap of α-spin (β-spin) states increases (decreases) as the external electric field increases. For a given external field, the band gap of α-spin (β-spin) increases (decreases) rapidly as the inter-ribbon distance increases. When we further increase the inter-ribbon distance, the band gap increases (decreases) slowly, and shows a saturation tendency. The wider the ribbon, the stronger the effect of Dx on the magnitude of the band gap. These observations can also be explained by the screening effect of the ribbon array. Another interesting result is that the semiconducting ZGNRs can be converted into half metals either by increasing the external field or by increasing the inter-ribbon distance. Under an external electric field of 0.2 V/A$^0$, the band gap of 8-ZGNR is decreased from 0.96 eV at Dx=5A$^0$ to 0.0 eV at Dx = 30 A$^0$. In addition, the band gap of 10-ZGNR is decreased from 0.67eV at Dx = 5 A$^0$ to 0.0eV at Dx = 20 A$^0$ with an external field of 0.2V/A$^0$.

Finally, we consider the screening effect on the magnetic moment of the edge carbon atoms. Figure 6 exhibits the magnetic moment on the edge carbon atoms for the A-sites and B-sites of ZGNRs under external electric fields of 0.0, 0.1 and 0.2 V/A$^0$. Without the external electric field, the total magnetic moment of ZGNR is zero, and the localized edge states on each edge atom have opposite and equal magnetic moment. For a given inter-ribbon distance, the magnetic moment increases as the width of the ribbon increases. This is also in good agreement with the previous calculation[41]. The inter-ribbon distance can slightly affect the edge magnetism. The magnetic moment on the edge atom decreases slowly and then tends toward saturation as the inter-ribbon distance increases due to the screening effect of the ribbon array.

When an external electric field is applied, the magnetic moment on each edge atom is suppressed. The modulation of the magnetic moment of the edge atom at the A-site is stronger than that of the edge atom at the B-site. Therefore, the total magnetic moment is nonzero under an external electric field. It is clear that the inter-ribbon distance can modulate the edge magnetism. The magnetic moment on the edge atom decreases slowly and then tends toward saturation as the inter-ribbon distance increases, which is also due to screening effect of the ribbon array. The wider the ribbon, the stronger the effect of the inter-ribbon distance on the edge magnetism.

Conclusions

In conclusion, we have systematically investigated the effect of inter-ribbon distance on the work function, field enhancement factor, band gap and edge magnetism of zigzag graphene nanoribbons using first-principles density functional theory. We found that the work function of ZGNRs increases rapidly as inter-ribbon distance Dx increases, which is caused by the positive dipole array at the edge of the ribbon. For a given Dx, the work function of ZGNRs decreases as the width of the ribbon increases. The wider the ribbon, stronger the effect of Dx on the work function. Using a simple linear model, we obtain the work function of ZGNR of any ribbon-width. The field enhancement increases rapidly when the inter-ribbon distance increases as long as Dx < W. When we further increase Dx, the field enhancement increases slowly and then tends toward saturation. The inter-ribbon distance of ZGNRs can modulate the magnitude of the band gap and magnetic moment of the edge atom. These observations can be explained by the screening effect of the ribbon arrays. Our findings not only provide an insight into understanding the screening effect on the characteristics of the ZGNRs, but also provide a guideline for their efficient application in field devices.

## Method of Calculation

The calculations were performed by the density functional theory (DFT)[42] and the local spin density approximation (LSDA)[43] using the Ceperley-Alder form of exchange-correlation functional[44] and highly accurate project augmented wave (PAW) method[45] with a plane wave cutoff of 400 eV, as implemented in the Vienna *ab-initio* simulation package (VASP)[46-48]. For Brillouin-zone integrations, the 1×80×1 *k*-points mesh with gamma centered grid is utilized for all systems studies. The atomic positions were relaxed until the magnitudes of the forces became less than 0.02 eV/Å.

Acknowledgments

The authors acknowledge the National Center for Theoretical Sciences in Taiwan and financial support from the Ministry of Science and Technology of Taiwan (MOST) under Grant Nos. MOST-105-2112M-194-006 (T. C. Leung) and MOST-104-2112-M-492-001 (W. S. Su). The authors are grateful to the National Center for High-Performance Computing (NCHC) for computer time and facilities.

Author contributions

T.C. Leung and W.S. Su conceived and initiated the study. H. Hu and, T.C. Lin performed first principles calculations. H. Hu, T.C. Lin, T.C. Leung, and W.S. Su performed the detailed analysis and contributed the discussions. T.C. Leung and W.S. Su wrote the manuscript. All authors reviewed the manuscript.

Additional information

Competing financial interests: The authors declare no competing financial interests.

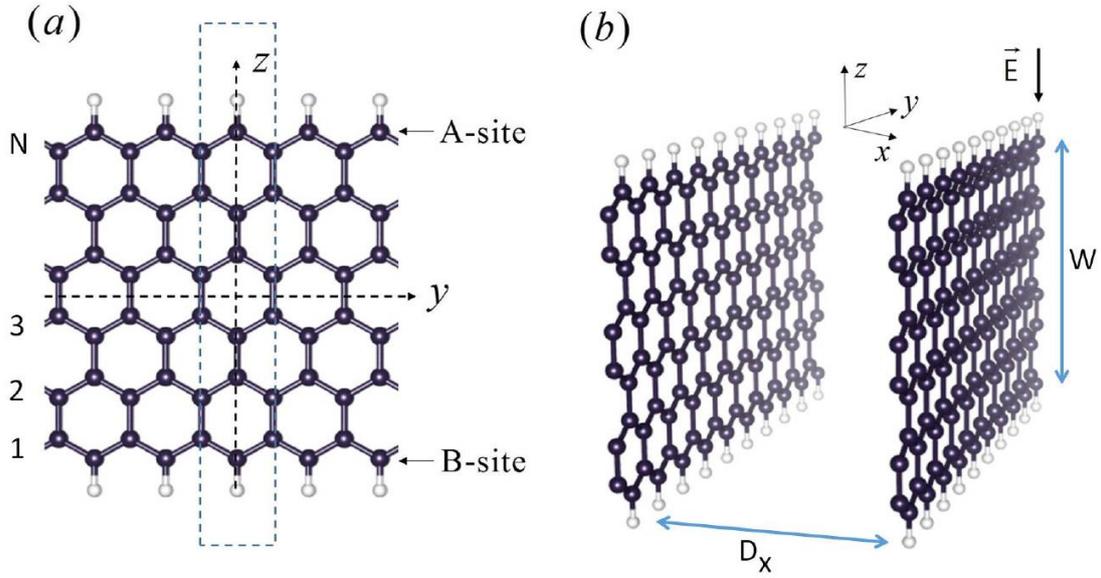

Fig. 1: (Color online) (a) Structural schematics and (b) orientation of the unit cell of a 6-ZGNR. Hydrogen atoms on the edges are denoted by small blue beads. The dashed rectangle indicates a unit cell. The ribbon width and inter-ribbon distance are denoted by W and Dx, respectively. The external electric field E is represented by an arrow in (b).

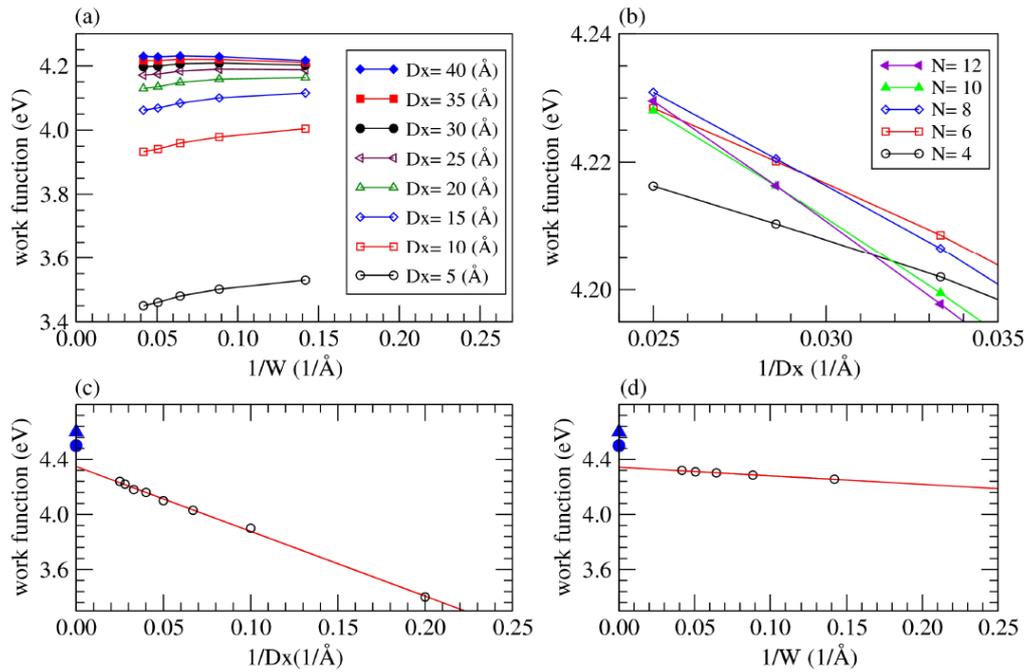

Fig. 2: (Color online) Work function of ZGNR, calculated by local density approximation (LDA), as a function of (a) inverse ribbon width $1/W$ and (b) inverse inter-ribbon distance $1/D_x$. The lines in (a) and (b) serve as visual guides. (c) Work function of wide ZGNRs, calculated by the linear fit to the results in (a), as a function of inverse inter-ribbon distance. (d) Work function of isolated ZGNR, calculated by the linear fit to the results in (b), as a function of inverse ribbon width. The lines in (c) and (d) are the linear fit to the data. The blue circle and triangle in (c) and (d) indicate the theoretical[38] and experimental[39] work function of graphene.

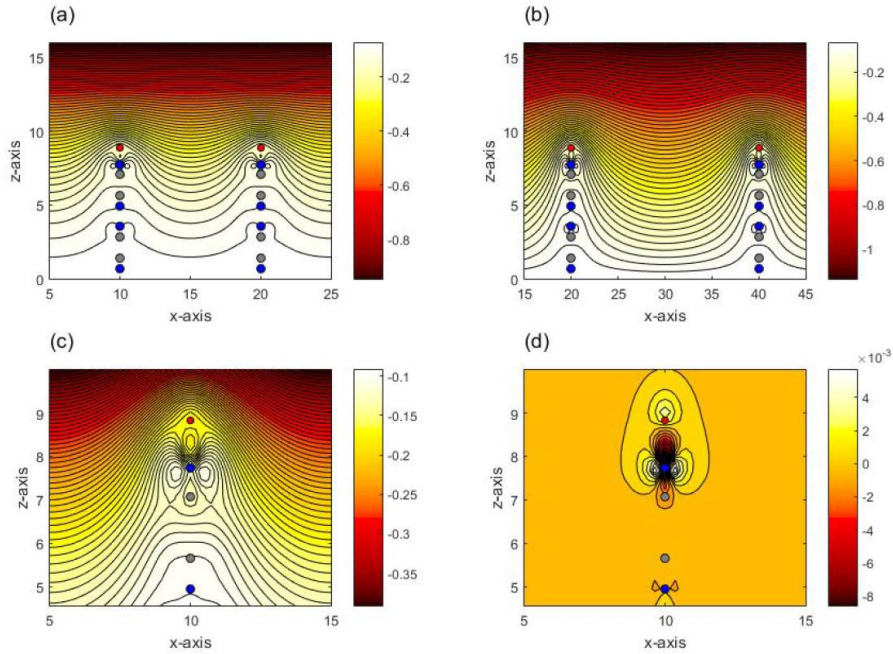

Fig. 3: (Color online) Space distribution of field-induced Coulomb potential around a 8-ZGNR viewed from the plane y=0 with inter-ribbon distances of (a) 10 $A^0$ and (b) 20 $A^0$. Field-induced Coulomb potential and field-induced charge density in the vicinity of the edge of a 8-ZGNR in the plane $y = 0$ with $Dx = 10$ $A^0$ are shown in (c) and (d), respectively. The red circles denote hydrogen atoms and blue circles represent carbon atoms in the plane $y = 0$. The gray circles represent carbon atoms in the plane $y = 1.22$ $A^0$.

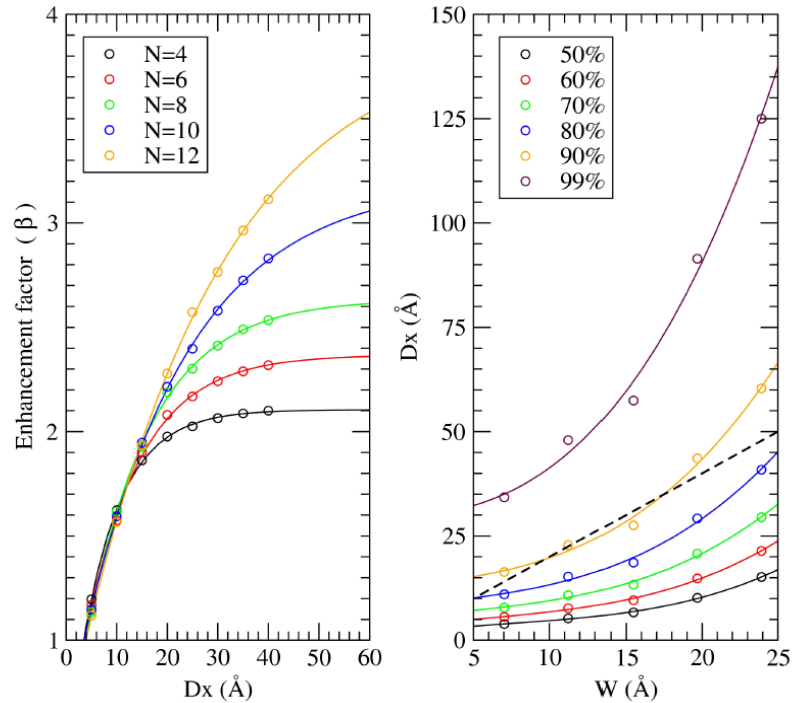

Fig. 4: (Color online) (a) Field enhancement factor β of N-ZGNR as a function of inter-ribbon distance under an electric field of E=0.1V/ $A^0$. (b) Inter-ribbon distance as a function of ribbon width such that the enhancement factor is 50, 60, 70, 80, 90 and 99% of its saturated value. The open diamonds in Fig. 4(a) and 4(b) are the field enhancement factor calculated by LDA. The solid lines in Fig. 4(a) denote the fitting curves calculated by Eq. (1). The solid lines in Fig. 4(b) denote the third degree polynomial curve-fitting results. The dashed line is given by Dx = 2 W.

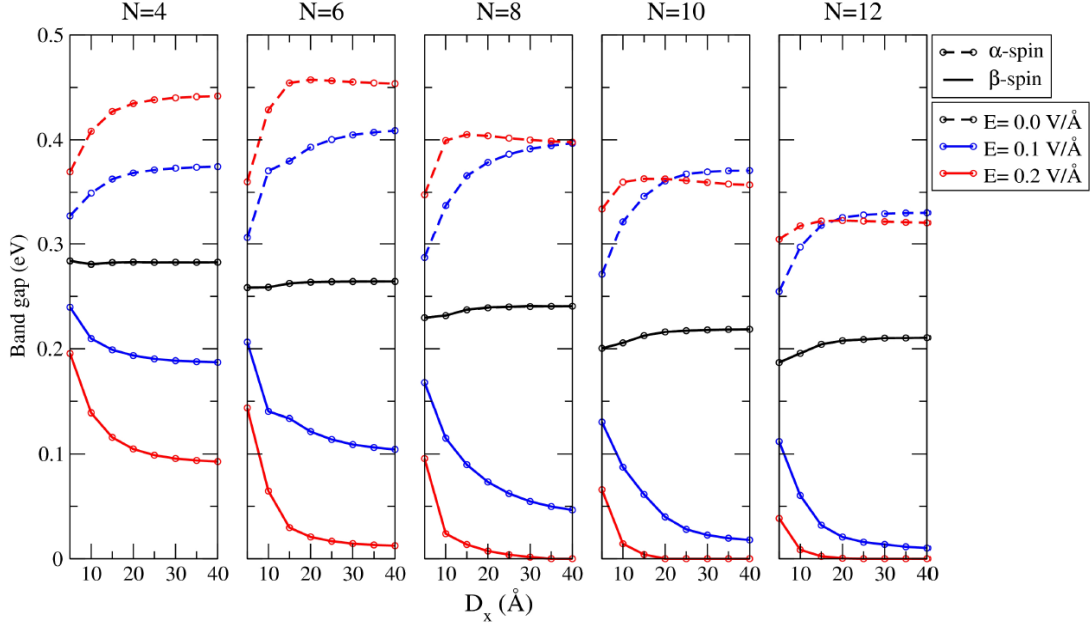

Fig. 5: (Color online) Band gap of N-ZGNRs as a function of inter-ribbon distance Dx under external electric field of 0.0, 0.1, and 0.2 V/A⁰. The band gap of α-spin and β-spin states are indicated by dashed and solid lines, respectively. Note that the α-spin and β-spin states are degenerate under zero external field.

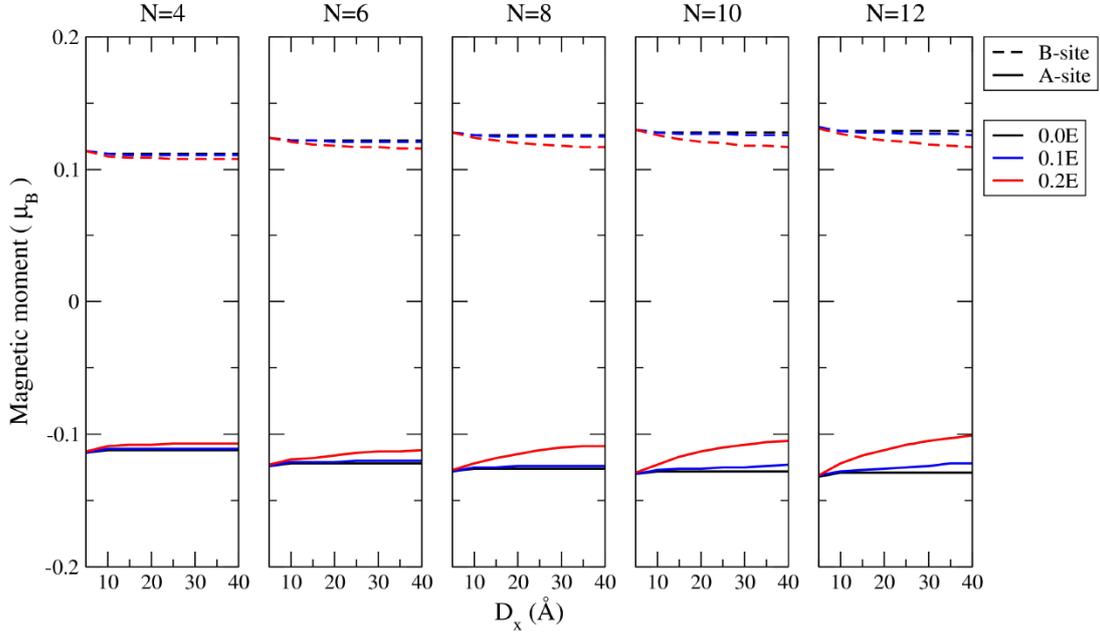

Fig. 6: (Color online) Magnetic moment of N-ZGNRs as a function of inter-ribbon distance Dx under external electric field of 0.0, 0.1, and 0.2 V/A⁰. The magnetic moment on the edge carbon atom for A-sites and B-sites are indicated by dashed and solid line, respectively.